\documentclass[conference]{IEEEtran}
\IEEEoverridecommandlockouts
\usepackage{cite}
\usepackage{amsmath,amssymb,amsfonts}
\usepackage{algorithmic}
\usepackage{graphicx}
\usepackage{textcomp}
\usepackage{xcolor}
\def\BibTeX{{\rm B\kern-.05em{\sc i\kern-.025em b}\kern-.08em
    T\kern-.1667em\lower.7ex\hbox{E}\kern-.125emX}}
\begin{document}

\title{CONVERGE: A Vision-Radio Research Infrastructure Towards 6G and Beyond}

\author{\IEEEauthorblockN{Filipe B. Teixeira\IEEEauthorrefmark{1}\IEEEauthorrefmark{2},
Manuel Ricardo\IEEEauthorrefmark{2}\IEEEauthorrefmark{1},
André Coelho\IEEEauthorrefmark{1}\IEEEauthorrefmark{2}, 
Hélder P. Oliveira\IEEEauthorrefmark{1}\IEEEauthorrefmark{3},
Paula Viana\IEEEauthorrefmark{4}\IEEEauthorrefmark{1},\\
Nuno Paulino\IEEEauthorrefmark{1}\IEEEauthorrefmark{2},
Helder Fontes\IEEEauthorrefmark{1}\IEEEauthorrefmark{2},
Paulo Marques\IEEEauthorrefmark{5},
Rui Campos\IEEEauthorrefmark{2}\IEEEauthorrefmark{1}, and
Luis M. Pessoa\IEEEauthorrefmark{1}\IEEEauthorrefmark{2}}
\IEEEauthorblockA{\IEEEauthorrefmark{1}INESC TEC, Porto, Portugal}
\IEEEauthorblockA{\IEEEauthorrefmark{2}Faculdade de Engenharia, Universidade do Porto, Portugal}
\IEEEauthorblockA{\IEEEauthorrefmark{3}Faculdade de Ciências, Universidade do Porto, Portugal}
\IEEEauthorblockA{\IEEEauthorrefmark{4}ISEP, Polytechnic of Porto, Portugal}
\IEEEauthorblockA{\IEEEauthorrefmark{5}Allbesmart Lda., Portugal}
\IEEEauthorblockA{\IEEEauthorrefmark{1}e-mail: filipe.b.teixeira@inesctec.pt}
\thanks{This work was supported by the CONVERGE project which has received funding under the European Union’s Horizon Europe research and innovation programme under Grant Agreement No 101094831.}  
}

\maketitle

\begin{abstract}
Telecommunications and computer vision have evolved separately so far. Yet, with the shift to sub-terahertz (sub-THz) and terahertz (THz) radio communications, there is an opportunity to explore computer vision technologies together with radio communications, considering the dependency of both technologies on Line of Sight. The combination of radio sensing and computer vision can address challenges such as obstructions and poor lighting. Also, machine learning algorithms, capable of processing multimodal data, play a crucial role in deriving insights from raw and low-level sensing data, offering a new level of abstraction that can enhance various applications and use cases such as beamforming and terminal handovers.

This paper introduces CONVERGE, a pioneering vision-radio paradigm that bridges this gap by leveraging Integrated Sensing and Communication (ISAC) to facilitate a dual ``View-to-Communicate, Communicate-to-View'' approach. CONVERGE offers tools that merge wireless communications and computer vision, establishing a novel Research Infrastructure (RI) that will be open to the scientific community and capable of providing open datasets. This new infrastructure will support future research in 6G and beyond concerning multiple verticals, such as telecommunications, automotive, manufacturing, media, and health.
\end{abstract}


\section{Introduction}
Integrated Sensing and Communication (ISAC) is emerging as a key trend for 6G, driven by the shift towards higher frequency bands and larger antenna arrays, which brings communications and sensing systems closer in terms of hardware and signal processing algorithms. This convergence is expected to enhance wireless networks with environment sensing capabilities and increase resource usage efficiency~\cite{Saa19}.

Large/Reconfigurable Intelligent Surfaces (LIS/RIS) will significantly contribute to ISAC, offering reconfigurable wireless environments and improving the performance of communications and sensing, particularly under non-line-of-sight conditions~\cite{Liu22}. The trend towards higher frequencies coupled with large antenna arrays and/or RIS enables greater communications capacity and spatial multiplexing, along with improved sensing accuracy, facilitating localization and device-free environment mapping and imaging of the surrounding environment, creating opportunities for computer vision technologies.

The convergence of telecommunications and computer vision, once distinct fields, brings up new opportunities for innovation, namely using computer vision to predict wireless channel dynamics and integrating radio-based sensing to boost computer vision applications. Research in this domain combines wireless communications, computer vision, sensing, and machine learning, enabling a wide range of innovative applications. Addressing this interdisciplinary challenge requires advanced Research Infrastructures (RI) and suitable tools.

\begin{figure}
  \centering
  \includegraphics[width=0.47\textwidth]{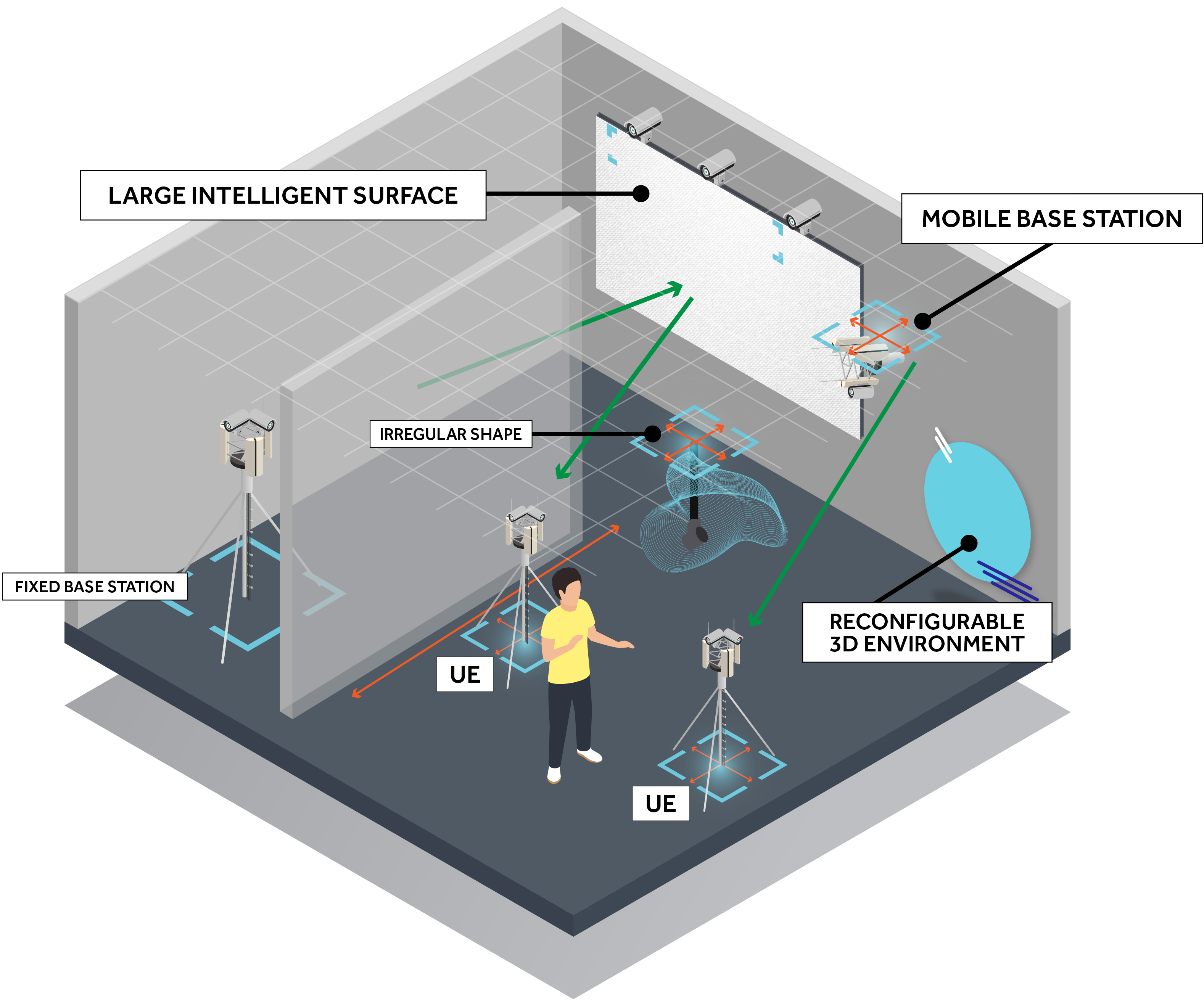}
  \caption{Illustration of the CONVERGE anechoic chamber.}
  \label{fig:CONVERGE_chamber}
\end{figure}

This paper introduces CONVERGE, a pioneering vision-radio paradigm that bridges this gap by leveraging ISAC to facilitate a dual ``View-to-Communicate, Communicate-to-View'' approach. Four major tools will be developed as part of the CONVERGE RI: 1) the vision-aided LIS, 2) the vision-aided fixed and mobile base station, 3) the vision-radio simulator and 3D environment modeler, and 4) the machine learning (ML) algorithms for multimodal data. These tools will integrate with the CONVERGE Chamber (Fig.~\ref{fig:CONVERGE_chamber}), enabling the collection of experimental data from radio communications, radio sensing, and vision sensing; the CONVERGE Simulator, where data collection will be possible using the CONVERGE Digital Twin; and the CONVERGE Core, responsible for the application function, session and data orchestration, time synchronization, open data repository, and ML algorithms. CONVERGE is aligned with the ESFRI SLICES-RI~\cite{Fdi2022}. It will serve as an RI that will provide the scientific community with open datasets of both experimental and simulated data. This framework enables research in 6G and beyond addressing various verticals, including telecommunications, automotive, manufacturing, media, and health.

The paper is structured as follows. Section II provides a review of the state of the art regarding infrastructures supporting experimentally-driven research. Section III presents the proposed vision-radio research infrastructure and tools. Section IV introduces the CONVERGE high-level, service-level, and gNB reference architectures. Section V presents CONVERGE use cases. Finally, Section VI and Section VII close the paper with the discussion and main conclusions.

\section{Review of the state of the art}
Experimental testbeds such as Bristol Is Open~\cite{bristol}, ADRENALINE~\cite{adrenalina}, GENI~\cite{geni}, Emulab~\cite{Emulab}, Fed4Fire~\cite{Fed4Fire}, ORBIT~\cite{ORBIT}, COSMOS~\cite{Cosmos}, and POWDER-RENEW ~\cite{Powder} are key resources for testing the next generation of communications, available to both academic and industrial researchers.

These platforms vary in focus, from smart city and IoT research (Bristol Is Open) to a circuit-switched optical testbed designed for large-scale optical transport networks and edge computing research 
(ADRENALINE). Fed4FIRE~\cite{Fed4Fire} focuses on creating a federation of testbeds for efficient utilization of experimental resources. They support experiments in diverse fields including energy, health, and transportation.

ORBIT~\cite{ORBIT} offers a platform for reproducible experimentation and realistic evaluation of protocols and applications. It features the Radio Grid Testbed, which utilizes a 20x20 two-dimensional grid of programmable radio nodes.

5GINFIRE~\cite{SILVA2019101895} aims to create an ecosystem for 5G Network Function Virtualization, integrating existing and new test facilities for software-driven architectural experimentation across various industries. TUAS~\cite{10.1007/978-3-319-76207-4_21} is working on a 5G ecosystem to support smart cities and other verticals, focusing on sub-6 GHz spectrum and exploring expansion frequencies.

COSMOS~\cite{Cosmos}, and POWDER-RENEW~\cite{Powder} focus on ultra-high bandwidth and low latency, and massive MIMO technologies, respectively. They demonstrate the potential of combining resources across testbeds to overcome limitations such as waiting times and experiment size constraints. COSMOS draws inspiration from networking and wireless testbeds such as GENI~\cite{geni}, Emulab~\cite{Emulab}, PlanetLab~\cite{PlanetLab}, OneLab~\cite{OneLab}, CloudLab~\cite{CloudLab}, and notably, ORBIT~\cite{ORBIT}. 

The purpose behind the accessibility of the testbeds is to facilitate the testing of prototypes without requiring the researchers to invest in a dedicated testing infrastructure. However, due to limited resources within these testbeds, there are often extended waiting periods for experiment approval, and the size of experiments becomes restricted. To overcome these limitations, inter-testbed experiments can be conducted such as in the case of COSMOS~\cite{Cosmos} and POWDER-RENEW~\cite{Powder}. By leveraging resources from multiple testbeds, researchers can create a more extensive and collective resource pool. 

Recent innovations explore integrating visual sensing with communications to enhance research, as seen in Bristol Is Open and COSMOS, where the utilization of visual sensing to enhance communications research by leveraging video frames for environmental perception has been demonstrated. Nonetheless, these visual sensors do not offer the same perspective as the communications stations, nor do they provide adequate trace resolution, repeatability, or reproducibility~\cite{Sharma}. The combination of video and radio is also missing on the ESFRI SLICES-RI~\cite{Fdi2022}. The potential for innovation in this combined research area is vast, not only due to the wide range of innovative applications it enables but also because of the existing expertise available in Europe in these domains. However, to fully assess the possibilities of this convergence, adequate RIs and tools need to be readily available.

\section{Proposed Vision-Radio \\Research Infrastructure}
The basis of the CONVERGE vision-radio research infrastructure is the CONVERGE anechoic Chamber, shown in Fig.~\ref{fig:CONVERGE_chamber}, which will enable the collection of experimental data from radio communications, radio sensing, and vision sensing. A CONVERGE Digital Twin of the chamber will be developed, allowing to carry experiments without access to the physical chamber.

\subsection{Vision-aided large intelligent surface}
RIS are crucial in smart radio environments~\cite{Ren19}, offering dynamic control over wireless channel to support a wide range of applications with minimal infrastructure. Large RIS (LIS) are anticipated to facilitate new applications aligned with ISAC like programmable holography and precise energy focusing, enhancing signal delivery and reducing inter-user interference. An RIS manipulates incident waves through a planar array of elements excited by an incident plane or spherical wavefront ~\cite{Ren20}, achieving targeted signal distribution with discretized phase control, often using positive–intrinsic–negative diodes.

Research on RIS has traditionally focused on the verification of pre-designed functions~\cite{Li23}. Next-generation (6G) advancements require the knowledge of user locations for high-directive links. Real-time computing and control of a self-adaptive RIS phase profile based on electromagnetic feedback from complex environments remains an open research topic. 

Computer Vision (CV) has progressed in complex tasks like object detection and tracking~\cite{Han18}, suggesting its utility in enhancing communications through environmental sensing. RIS-based sensing, especially at frequencies below 6 GHz can, in turn, aid CV applications.  

We introduce the vision-aided LIS (VA-LIS), integrating a large programmable RIS with cameras to capture multiple video perspectives. Designed for controlled room experiments, VA-LIS works alongside advanced communications tools such as 5G Base Station and massive MIMO,  high precision 3D positioning, and environment sensing, including human sensing and microwave holography.

\subsection{Vision-aided fixed and mobile base station}
The CONVERGE Chamber will incorporate both fixed and mobile vision-aided base stations to facilitate communications and experiments with User Equipment (UE) and illuminate the VA-LIS. The fixed base station offers static networking conditions, but with relocatable positioning in the CONVERGE Chamber between different experiments, while the mobile base station introduces dynamic positioning via an overhead crane system, enabling an obstacle-aware mobile communications cell adding the possibility of simulating Unmanned Aerial Vehicles (UAVs) communications with multiple degrees of freedom in a safe and predictable environment. 

These stations will support emerging 6G use cases at mmWave frequencies, incorporating beamforming, opportunistic scheduling, and multi-access UE capabilities. Open Radio Access Network (O-RAN) interfaces and xApps enhances communications performance and time-stamped and space-referenced RF sensing and communications traces \cite{o-ran-org}, while integration with video cameras and LiDAR, in cooperation with LIS, provides a comprehensive video and RF-based perspectives of the environment. A vision-aided mobile UE will further enrich data exchange and multimodal characterization, supported by OpenAirInterface software \cite{oai} and Software-Defined Radio (SDR) for enhanced flexibility. Both station types are designed for continuous operation, connected to a fiber-optic network for high-speed backhaul.

\subsection{Vision-radio simulator, 3D environment modeler and Network Simulator} 
Traditional radio signal prediction methods, including ray tracing, lack accuracy in various scenarios due to their need for complex 3D models \cite{Liu2021}. The vision-radio simulator and 3D environment modeler tool addresses this by creating a Digital Twin of a real anechoic chamber, comprising: 1) a 3D Environment Modeler (3D-S), 2) a Vision-Radio Simulator (VR-S), and 3) a Network Simulator (NET-S). This tool enables digital replication of environments, simulates radio signal propagation using geometric and electromagnetic models, and generates simulated traffic based on realistic channel models. Data from LiDAR, video cameras, and radio devices in the real anechoic chamber enrich the simulations, offering a comprehensive tool for experiment reproducibility, repeatability, and scalability.

This digital twin facility allows researchers to conduct experiments remotely, enhancing the design and optimization of algorithms, including ML-based ones, before final validation in the actual chamber. It supports complex signal propagation studies and the development of emerging wireless technologies by simulating diverse and complex communications scenarios.

\subsection{Machine learning algorithms for multimodal data}
Machine learning (ML) effectively advances learning from data to implement future tasks, leveraging knowledge from existing datasets. In wireless communications, ML improves signal recognition, spectrum sensing, and waveform design tasks \cite{AML-WC}. RIS are noted for their potential in boosting network capacity and coverage, especially when using high-frequency waves where obstruction from objects will have a great impact. ML's role extends to enhancing resource and energy management, security, beamforming, and channel estimation \cite{ML-RIS}. In multimedia content analysis, ML outperforms traditional computer vision techniques by offering superior object detection and identification, face recognition, and image segmentation \cite{ML-CV}. The shift towards analyzing multimedia, including video, audio, and text, through a multimodal approach has shown better performance than individual data analysis methodologies \cite{DL-MM}.

The CONVERGE architecture integrates the processing of heterogeneous data, including video, LiDAR, RF sensing signals, and timestamped and space-referenced communications traces, including received power, signal-to-noise ratio, antenna radiation patterns, and object positioning, facilitating comprehensive scenario analysis. It supports the adaptation and creation and fine-tuning of existing ML models to meet specific research needs, performing data anonymization and synchronization. This architecture not only enables the exploration of various data fusion techniques and evaluation metrics but also the development of innovative data visualization methods. A key feature is its API, designed to foster interaction with other CONVERGE components and external user access, enhancing the utility and applicability of ML in diverse applications.

\section{CONVERGE Reference Architecture}
\begin{figure*}
  \centering
  \includegraphics[width=0.77\textwidth]{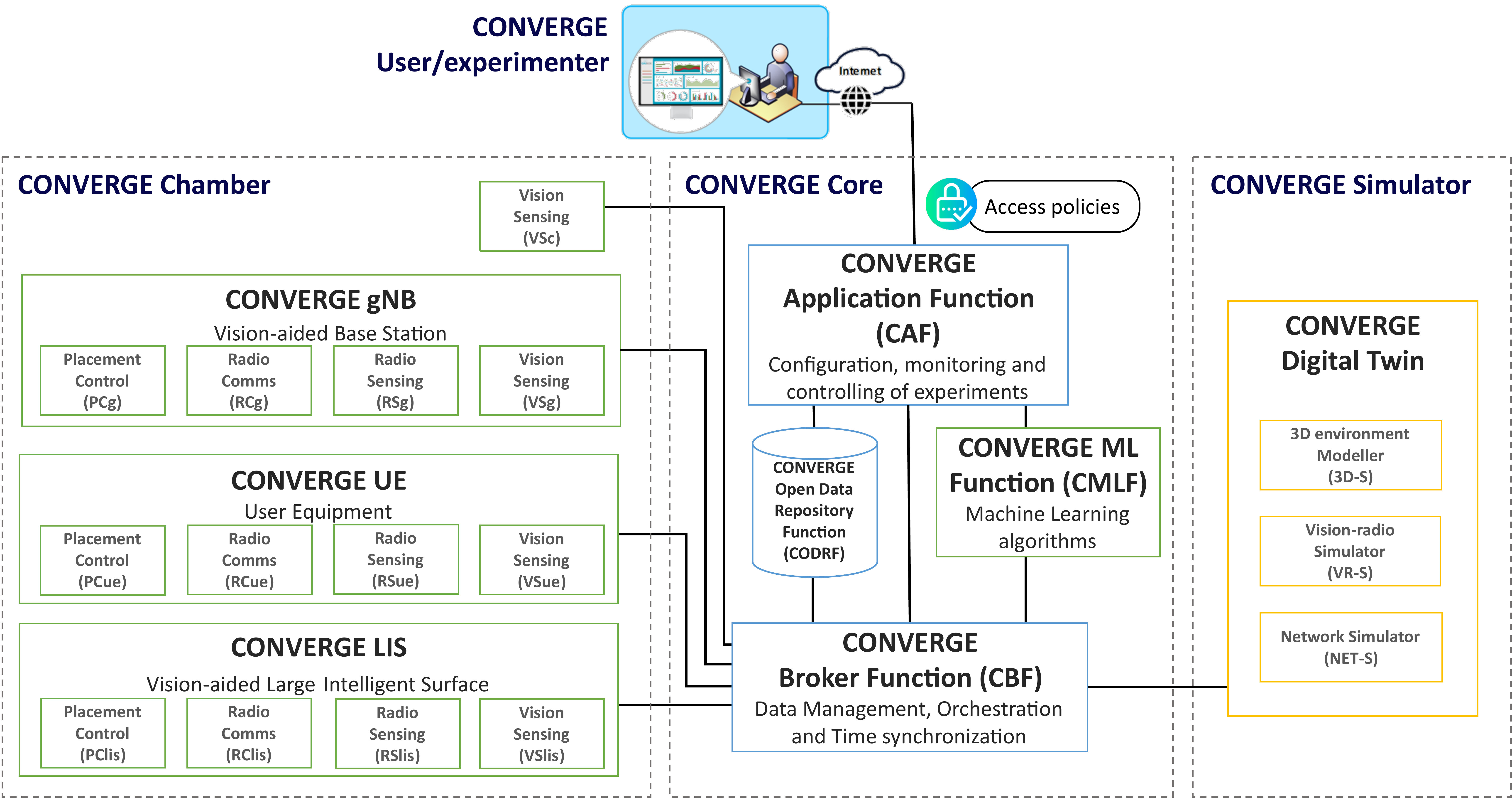}
  \caption{CONVERGE high-level architecture.}
  \label{fig:high-levelarchitecture}
\end{figure*}

\begin{figure*}
  \centering
  \includegraphics[width=0.77\textwidth]{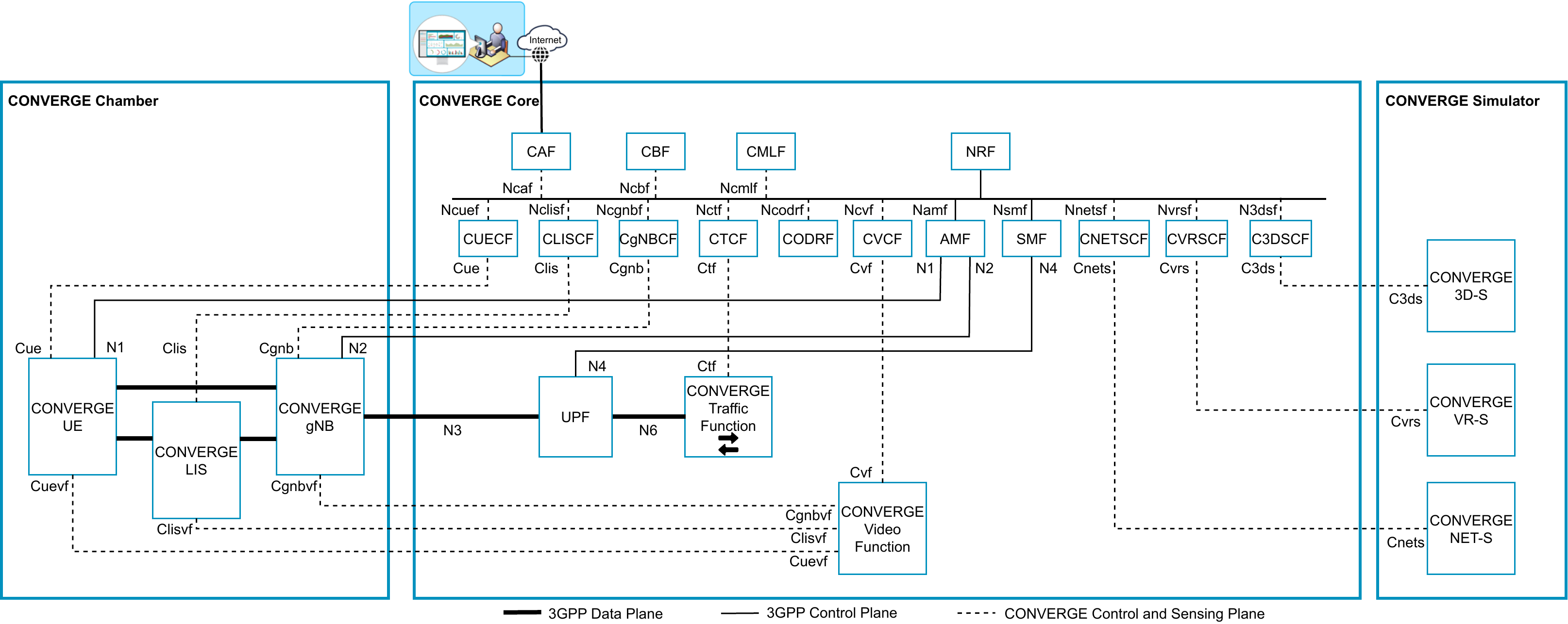}
  \caption{CONVERGE service-oriented architecture.}
  \label{fig:service-orientedarchitecture}
\end{figure*}

The CONVERGE reference architecture is depicted in Fig.~\ref{fig:high-levelarchitecture} and Fig.~\ref{fig:service-orientedarchitecture}. The former introduces the key building blocks of the CONVERGE system;  the latter decomposes these blocks, presents their control as virtual network functions, and identifies the interfaces. Finally, we present the gNB architecture.

\subsection{High-level architecture}
The CONVERGE architecture shown in Fig. \ref{fig:high-levelarchitecture} integrates three key components for comprehensive experimentation: the CONVERGE Chamber for physical experiments with gNB, UE, and LIS equipment in different research infrastructures, including mobility scenarios; the CONVERGE Simulator for scenario planning and testing via a Digital Twin; and the CONVERGE Core for managing experiments, data, and ML models. This setup supports both physical and simulated studies, streamlining equipment mobility, data interfacing, and management. The CONVERGE infrastructure allows external users to conduct real-time physical or off-line virtual experiments, leveraging the chamber, the simulator, ML tools, and datasets for advanced research.

The CONVERGE Chamber is equipped with a gNB, a UE, and a LIS, each featuring controls for placement: Placement Control of gNB (PCg), Placement Control of the UE (PCue), Placement Control of LIS (PClis); radio communications defined for 5G: Radio Communications of gNB (RCg), Radio Communications of UE (RCue), Radio Communications of the LIS (RClis); radio sensing: Radio Sensing of gNB (RSg), Radio Sensing of the UE (RSue), Radio Sensing of LIS (RSlis), and vision sensing functionalities: Video Sensing of gNB (VSg), Video Sensing of UE (VSue), Video Sensing of LIS (VSlis). An optional video sensing of chamber activities can exist; Visdeo Sensing of Chamber (VSc). These components are coordinated by the CONVERGE Core, which facilitates experiment control and data storage.

For remote access for radio and image task, FPGA-based Systems-on-Chip such as AMD's RF-SoC family can be used, providing versatile interfaces and support for Python, simplifying setup and enabling real-time ML applications and data visualization through Jupyter servers \cite{Goldsmith2020}.

The CONVERGE Core oversees equipment operations, enabling user interaction for experiment setup, monitoring, and control. It includes the CONVERGE Application Function (CAF) for user interface and access policies, the CONVERGE Broker Function (CBF) for session orchestration inside the chamber, the CONVERGE Open Data Repository Function (CODRF) for data storage, including open datasets, and the CONVERGE Machine Learning Function (CMLF) for accessing ML tools. These tools are accessible to external users, CONVERGE equipment, or CONVERGE Simulator tools.

The CONVERGE Simulator, with its 3D Environment Modeller (3D-S), Vision Radio Simulator (VR-S), and Network Simulator (NET-S), creates a Digital Twin for recreating digitally a physical environment, modelling the radio and vision communications in the environment by using ray tracing-like capabilities, and generating simulated traffic considering the realistic physical channel models obtained by the previous two tools. The simulation sessions are controlled by the CBF, storing the simulator data in the CODRF.

\subsection{Service-oriented architecture}
The CONVERGE architecture adopts a service-oriented design aligned with the 5G core network (Fig. \ref{fig:service-orientedarchitecture}), based on 3GPP standards \cite{3gpp}. This alignment is chosen for several reasons: 1) it leverages essential 5G core functionalities such as Access and Mobility Management Function (AMF), Session Management Function (SMF), and User Plane Function (UPF) necessary for UE and gNB operations, 2) utilizes the comprehensive 5G NR specifications, and 3) facilitates the integration of new functions for controlling the CONVERGE Chamber and Simulator. 

In the CONVERGE control and sensing plane (Fig.~\ref{fig:service-orientedarchitecture}, CONVERGE Core) the functionality to accommodate traffic servers is added (CONVERGE Traffic Function) as well as servers for receiving the radio sensing and vision information gathered from the CONVERGE equipment (CONVERGE Video Function). A set of new functions is added which will enable the simple and direct control of the CONVERGE Chamber equipment and CONVERGE Simulator tools. These control functions (CUECF, CLISCF, CgNBCF, CTCF, CODRF, C3DSCF, CVRSCF, and CNETSCF) will be implemented as web services (RESTful API) for easy and direct operation. These functions cover a range of control and data processing needs, with detailed specifications and interfaces outlined in the document "D1.2: Specification of interfaces and access policies (initial) \cite{CONVERGE}."

\subsection{gNB architecture}
\begin{figure}[t!]
  \centering
  \includegraphics[width=0.48\textwidth]{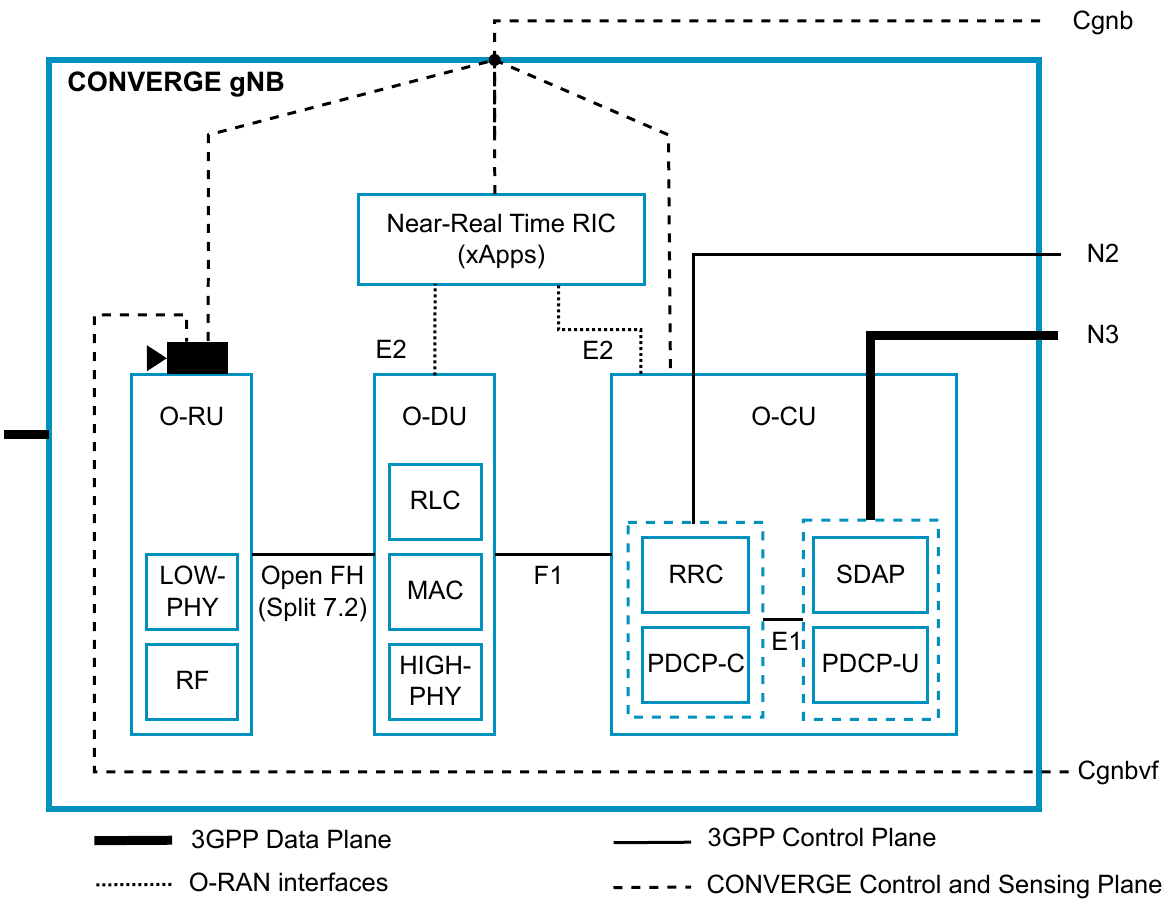}
  \caption{CONVERGE gNB architecture.}
  \label{fig:gnb-orientedarchitecture}
\end{figure}

The vision-aided gNB is a crucial part of the CONVERGE architecture. It is based on the O-RAN architecture, including its framework for supporting intelligent controllers, as shown in Fig.~\ref{fig:gnb-orientedarchitecture}.  O-RAN is particularly focused on the Radio Access Network (RAN), promoting interoperability and flexibility in 5G and future networks. The gNB integrates O-RAN's key components: Radio Unit (O-RU), Distributed Unit (O-DU), and Centralized Unit (O-CU), following the 3GPP F1 interface and functional split 7.2 \cite{3gpp}. Vision-sensing capabilities are added via a video camera on top of the O-RU in Fig.~\ref{fig:gnb-orientedarchitecture}. The CONVERGE gNB interfaces with the CONVERGE Core through a web service-based interface for remote access and control (Cgnb). This setup will support the development and validation of new xApps, software applications that run on the RAN Intelligent Controller (RIC) to improve network operations by leveraging both radio and vision data for near-real-time control and optimization in 5G networks.

\section{CONVERGE example use cases}
\begin{figure}[t!]
  \centering
  \includegraphics[width=0.49\textwidth]{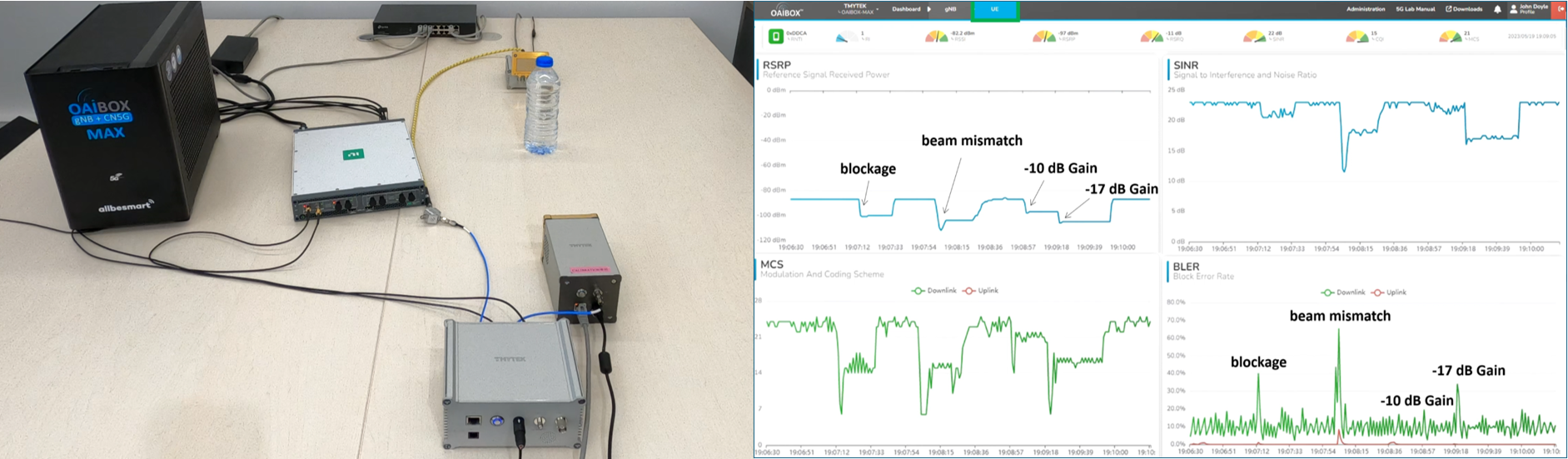}
  \caption{Initial setup for end-to-end 5G FR2 SA experimentation of the vision-aided proactive beam-switching mechanism.}
  \label{fig:usecase}
\end{figure}

The CONVERGE use case for proactive vision-aided beam-switching (telecommunications vertical)~\cite{CONVERGE} allows the simultaneous use of three tools: Vision-aided gNB, LIS, and ML algorithms. This setup tests a mechanism where beam failure, caused by sudden Line-of-Sight (LoS) blockages affecting UE channel conditions, is addressed by preemptively switching the beam through a beam-sweeping process. Traditionally, radio systems handle this through probing, causing delays and temporary 5G data link performance issues. CONVERGE utilizes visual data to capture environment features, such as object locations and their mobility pattern, anticipating bad channel conditions. A machine-learning algorithm can predict LoS signal blockage and trigger a vision-aided proactive beam switching. This approach is enhanced by LIS, where the gNB preemptively switches the beam to an LIS to reflect signals to the UE. An initial setup for end-to-end 5G FR2 SA experimentation can be seen in Figure \ref{fig:usecase}, based on OpenAirInterface (OAI) 5G FR2 stack with a NI USRP X410 for baseband processing, Up/Down mmWave converters and beamformers from TMYTEK and an OAIBOX~\cite{OAIBOX}. An intuitive web-based dashboard works as an abstraction layer to facilitate the monitoring and controlling of the 5G FR2 SA testbed.

The CONVERGE use case for image and RF sensing patient monitoring (health vertical)~\cite{CONVERGE} aims to infer a patient's posture and gait with the aid of RF sensing. Radio signals and patient posture are filmed while the patient moves through the camera to create a virtual patient for gait analyses. Then, ML algorithms will fuse this data to improve gait detection.

\section{Discussion}
CONVERGE introduces a new approach that combines radio sensing, vision, and communications, with a focus on testing through specialized tools and infrastructure, aligned with a 5G network architecture. However, challenges arise with real-time operations, orchestration of the different functions presented in the service-oriented architecture and synchronization of radio sensing and video sensing. Real-time machine learning (ML) inference, necessary for processing multiple streams and making immediate decisions like UE beam tracking, demands high computing power and low network latency. Current efforts are now focused on specifying and building each tool, designing the CAF, and data collection and distribution among different CONVERGE functions.

\section{Conclusions}
Telecommunications and computer vision are converging, with significant potential for innovation by leveraging visual data to predict wireless channel dynamics and enhancing computer vision applications through radio-based imaging. This interdisciplinary approach, integrating wireless communications, computer vision, sensing, and ML, opens up various innovative applications. 

This paper introduced CONVERGE, a pioneering vision-radio paradigm that bridges this gap by leveraging ISAC to facilitate a dual ``View-to-Communicate, Communicate-to-View'' approach. The tools developed within the RI were presented: 1) the vision-aided LIS, 2) the vision-aided base stations, 3) the vision-radio simulator with 3D modeling, and 4) the ML algorithms for processing multimodal data. This paper outlined high-level, service-oriented, and gNB architectures. Set to launch in 2026, CONVERGE will offer datasets for research targeting 6G and beyond verticals and aligning with ESFRI SLICES-RI. Future expansions aim at outdoor environments. Future work includes the expansion of CONVERGE for outdoor environments.

\bibliographystyle{IEEEtran}
\bibliography{references}

\end{document}